\documentclass[prc,letterpaper,twocolumn,showpacs,showkeys,lengthcheck,
               floatfix,nofootinbib,preprintnumbers,superscriptaddress]{revtex4-1} 

\usepackage{mathtools}
\usepackage{amssymb,amsfonts}
\usepackage{bm}
\usepackage{slashed}

\usepackage[svgnames]{xcolor}
\usepackage[english]{babel}
\usepackage{blindtext}
\usepackage{microtype}
\usepackage{tikz}
\usepackage{dcolumn}
\usepackage{multirow}
\usepackage[]{units}

\usepackage{graphicx}
\usepackage[caption=false]{subfig}

\usepackage[colorlinks=true,urlcolor=blue]{hyperref}
\usepackage{ifpdf}

\def\a{\alpha}
\def\b{\beta}

\def\g{\gamma}

\def\ve{\varepsilon}

\def\s{\sigma}
\def\o{\omega}

\def\G{\Gamma}
\def\L{\Lambda}

\def\S{\Sigma}

\def\pl{\partial}
\def\hs{\hspace}

\def\ol{\overline}
\def\no{\nonumber}

\def\lf{\left}
\def\rg{\right}

\newcommand{\sh}[1]{\slashed{#1}}

\def\calQ{\mathcal{Q}}

\def\be{\begin{equation}}
\def\ee{\end{equation}}

\newcommand{\Lagr}{\mathop{\mathcal{L}}}

\begin{document}

\title{Rho meson form factors in a confining Nambu--Jona-Lasinio model}

\author{Manuel~E.~Carrillo-Serrano}
\affiliation{CSSM and ARC Centre of Excellence for Particle Physics at the Tera-scale,\\
Department of Physics,
University of Adelaide, Adelaide SA 5005, Australia
}

\author{Wolfgang~Bentz}
\affiliation{Department of Physics, School of Science, Tokai University,
             4-1-1 Kitakaname, Hiratsuka-shi, Kanagawa 259-1292, Japan}

\author{Ian~C.~Clo\"et}
\affiliation{Physics Division, Argonne National Laboratory, Argonne, Illinois 60439, USA
}

\author{Anthony~W.~Thomas}
\affiliation{CSSM and ARC Centre of Excellence for Particle Physics at the Tera-scale,\\
Department of Physics,
University of Adelaide, Adelaide SA 5005, Australia
}

\begin{abstract}
Elastic electromagnetic form factors for the $\rho^+$ meson are calculated in a
Nambu--Jona-Lasinio model which incorporates quark confinement through the use of the proper-time regularization scheme.
A comparison is made with recent lattice QCD results and previous quark model calculations for 
static quantities and the Sachs form factors. The results are qualitatively in good agreement
with the lattice QCD calculations, with the exception of the quadrupole moment 
and corresponding form factor, which may be related to a lack of spherical symmetry on the lattice.
\end{abstract}

\pacs{
12.38.Aw,     
12.39.Fe,     
13.40.Gp,     
14.40.Be      
}

\maketitle
\section{Introduction \label{sec:intro}}

The structure of hadrons presents a remarkable challenge to the theory of
strong interactions -- quantum chromodynamics (QCD) -- and a critical feature of a hadron's structure is its
distribution of charge and magnetization, which is empirically related
to its electromagnetic form factors~\cite{Thomas:2001kw}. The
direct calculation of hadron form factors using QCD is currently only possible 
through lattice QCD, albeit limited to the low to moderate $Q^2$ region.
However, to gain insight into the relevant dynamical mechanisms behind the 
observed structure it is useful to work with models that approximate key
features of QCD. An important focus for this comparison are the 
meson form factors. Because of their short lifetimes~\cite{Agashe:2014kda} they present a unique 
challenge experimentally -- making both lattice QCD and model calculations critical. The pion form factor has
been successfully measured over a wide range of four momentum transfer, while
the vector meson form factors have not had the same amount of experimental exploration.
However, the BABAR collaboration has measured the cross-section for the 
reaction $e^{+} + e^{-} \to \rho^{+} + \rho^{-}$~\cite{Aubert:2008gm}, which has been
analyzed to garner information on the $\rho$-meson form factors~\cite{Dbeyssi:2011ep}.

The $\rho$ form factors, or equivalently the polarization amplitudes, have been calculated using
a variety of methods,
for example, phenomenological models~\cite{Adamuscin:2007dt,Gudino:2013jaa}, constituent 
quark models in the light front framework~\cite{Chung:1988my,Cardarelli:1994yq,deMelo:1997hh,Choi:1997iq,
Melikhov:2001pm,Jaus:2002sv,Choi:2004ww,Biernat:2014dea}, QCD sum rules~\cite{Aliev:2003ba,Samsonov:2003hs,Braguta:2004kx,Aliev:2004uj} and 
the Dyson-Schwinger equations~\cite{Hawes:1998bz,Bhagwat:2006pu,Roberts:2011wy,Pitschmann:2012by}.
The first attempts to compute $\rho$ form factors using lattice QCD were reported in Refs.~\cite{Andersen:1996qb,Hedditch:2007ex} 
in the quenched framework. The recent work of Owen \textit{et al.}~\cite{Owen:2015gva} 
and Shultz \textit{et al.}~\cite{Shultz:2015pfa} give two independent lattice QCD 
calculations based upon different approaches. These lattice results, and the previous work
with quark models, provides a solid background for comparison with results
computed within other models.

In this work we extend the $\rho$-meson form factor calculation of Ref.~\cite{Cloet:2014rja},
where the focus was a comparison with the axialvector diquark form factors which formed
a critical part of a nucleon form factor calculation. Here we use the same 
confining version of the Nambu--Jona-Lasinio (NJL) model~\cite{Nambu:1961tp,Nambu:1961fr}
to investigate the 
quark mass dependence of the $\rho$ form factors, and perform a detailed comparison 
with the lattice QCD results of Refs.~\cite{Owen:2015gva,Shultz:2015pfa}. 
Following Ref.~\cite{Cloet:2014rja} we include the dressing of the quark-photon vertex 
from the inhomogenerous Bethe-Salpeter equation and a pion cloud, which are critical for a good agreement with 
lattice results. Similar finding were made in Ref.~\cite{Ninomiya:2014kja}, where the
same framework was applied to the $\pi$ and $K$ form factors.

The outline of the paper is as follows: In Sec.~\ref{sec:NJL} we briefly review the 
NJL model as applied to $\bar{q}q$ bound states and 
the calculation of the $\rho$ electromagnetic form factors is discussed in Sec.~\ref{sec:FFs}. 
The results are compared to those from lattice QCD and various quark models in Sec.~\ref{sec:results}
and Sec.~\ref{sec:con} presents our conclusions.

\section{Nambu--Jona-Lasinio Model \label{sec:NJL}}

In its original formulation the NJL model successfully encapsulated the effects of dynamical chiral
symmetry breaking, where the pion emerged as a Goldstone boson and the nucleon was the fundamental
degree of freedom~\cite{Nambu:1961tp,Nambu:1961fr}. The NJL model has subsequently been
re-expressed with quarks as the fundamental constituents, making the relation with QCD evident. 
Importantly, the NJL model~\cite{Klevansky:1992qe} preserves the fundamental symmetries of QCD. In particular,
the generation of mass through the dynamical breaking of chiral symmetry is
beautifully illustrated. In contrast, quark confinement is not automatically incorporated in the model. However, it has been 
shown that it can be mimicked by the introduction of an infrared cutoff in the 
proper-time regularization scheme~\cite{Ebert:1996vx,Hellstern:1997nv,Bentz:2001vc}.
The NJL model has a history of success in the description of numerous meson~\cite{Klevansky:1992qe,Hatsuda:1994pi,Vogl:1991qt} and baryon~\cite{Hatsuda:1994pi,Vogl:1991qt} 
properties, including the nucleon parton distribution functions~\cite{Cloet:2005pp,Cloet:2005rt,Cloet:2007em,Bentz:2007zs,Matevosyan:2011vj} and 
electromagnetic form factors~\cite{Cloet:2014rja}. More recently these studies have been extended to 
the computation of the axial charges for strangeness conserved $\beta$-decays in the 
baryon octet~\cite{Carrillo-Serrano:2014zta} 
and possible insights into the solutions of long time enigmas in QCD, such as the $\Delta I = 1/2$ rule in 
kaon decays~\cite{Liu:2014vha}. It is this wealth of achievement, together with the recent developments in lattice QCD that 
encourage us to test whether the model gives an accurate description of $\rho$-meson properties.
  
In the application of the NJL model to the solution of the form factors of the $\rho$-meson,
we use a two-flavor NJL Lagrangian which in the $\bar{q}q$ interaction channel reads:
\begin{align}
\Lagr &= \bar{\psi}\lf(i\sh{\pl} - \hat{m}\rg)\psi \no \\
&+\tfrac{1}{2}\,G_{\pi}\lf[\lf(\bar{\psi}\psi\rg)^2 - \lf(\bar{\psi}\g_5\vec{\tau}\psi\rg)^2\rg] - \tfrac{1}{2}\,G_{\o}\lf(\bar{\psi}\g^{\mu}\psi\rg)^2 \no \\
&-\tfrac{1}{2}\,G_{\rho}\lf[\lf(\bar{\psi}\g^{\mu}\vec{\tau}\psi\rg) + \lf(\bar{\psi}\g^{\mu}\g_5\vec{\tau}\psi\rg)^2\rg],
\label{eq:lagrangian}
\end{align}
where $\vec{\tau}$ are the Pauli matrices representing isospin and $\hat{m} = \text{diag}\lf[m_u,\,m_d\rg]$
is the current quark mass matrix. We assume $m_u = m_d = m$. The fermion coupling $G_\pi$ represents the strength 
of the scalar ($\bar{q}q$) and pseudoscalar ($\bar{q}\g_5q$) interaction channels and
is responsible for the dynamical generation of the dressed quark masses through the breaking of 
chiral symmetry.  The strength of the vector-isoscalar and vector-isovector four fermion interactions is
given by $G_\o$ and $G_\rho$, respectively. The explicit breaking of $U(1)$ axial symmetry is often modeled by the inclusion
of an extra six-fermion determinant interaction term, which describes
the $\eta$ and $\eta'$ mass splitting~\cite{Klevansky:1992qe},
however, this is not directly related to our calculation so we do not consider it. We 
regularize the NJL interaction through the proper-time regularization scheme,
using an infrared cutoff ($\L_{IR}$) to remove unphysical decay thresholds
for hadrons into quarks~\cite{Ebert:1996vx,Hellstern:1997nv,Bentz:2001vc}. 

\begin{figure}[tbp]
\centering\includegraphics[width=\columnwidth,clip=true,angle=0]{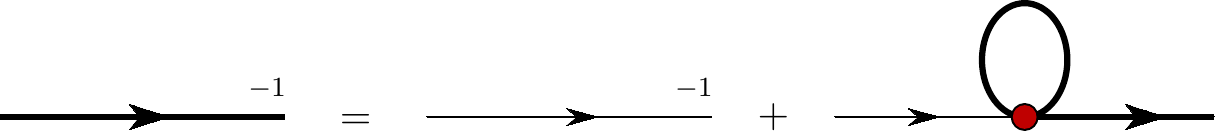}
\caption{(Colour online) The NJL gap equation in the Hartree-Fock approximation, where the thin line represents the elementary quark propagator, $S_{0}^{-1}(k) = \sh{k} - m + i\ve$, and the shaded circle represents the 4-fermion interaction.}
\label{fig:gapequation}
\end{figure}

The dressed quark masses are given by the solution of the gap equation
depicted in Fig.~\ref{fig:gapequation}, which in the proper-time scheme reads
\begin{align}
M = m + \frac{3}{\pi^{2}}\,M\,G_{\pi}\int_{1/\L_{UV}^{2}}^{1/\L_{IR}^{2}} d\tau\, \frac{e^{-\tau M^2}}{\tau^2},
\label{Eq:gap}
\end{align}
giving a dressed quark propagator of the form:
\begin{align}
S(k)^{-1} = \sh{k} - M  + i\ve.
\label{Eq:quarkpropagator}
\end{align}

\begin{figure}[tbp]
\centering\includegraphics[width=\columnwidth,clip=true,angle=0]{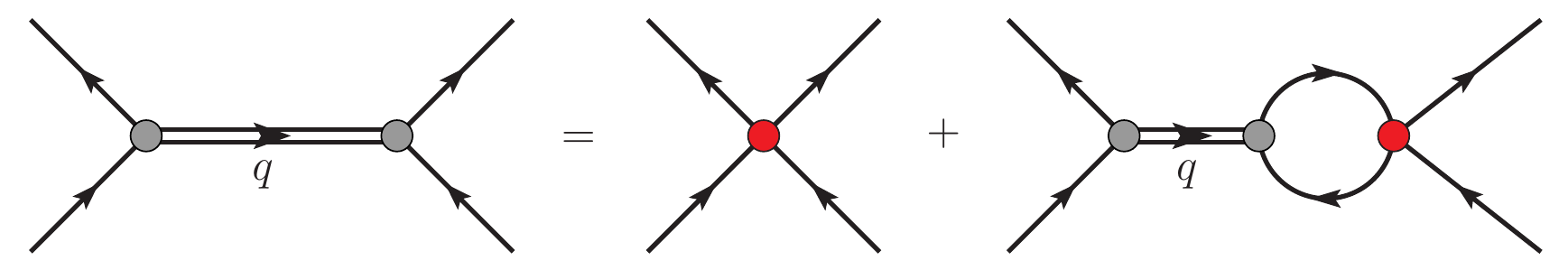}
\caption{(Colour online) Bethe-Salpeter equation for antiquark--quark (meson) correlations in the
NJL model using the random phase approximation.}
\label{fig:bethesalpetermesons}
\end{figure}

The description of mesons as $\bar{q}q$ bound states in the NJL model is obtained via the
Bethe-Salpeter equation (BSE) in the random-phase approximation, as illustrated in Fig.~\ref{fig:bethesalpetermesons}.
The solution of the BSE in each meson channel is given by a two-body $t$-matrix that 
depends on the nature of the interaction channel~\cite{Cloet:2014rja}, 
where the reduced $t$-matrices for the $\pi$, $\rho$ and $\omega$ mesons read
\begin{align}
\label{eq:tpion}
\tau_\pi(q) &= \frac{-2i\,G_\pi}{1 + 2\,G_\pi\,\Pi_{PP}(q^2)}, \\[0.7ex]
\label{eq:trho}
\tau^{\mu\nu}_{\rho(\o)}(q) 
&= \frac{-2i\,G_{\rho(\o)}}{1+2\,G_{\rho(\o)}\,\Pi_{VV}(q^2)} \no \\
&\hs{13mm} \times
\lf[g^{\mu\nu} + 2\,G_{\rho(\o)}\,\Pi_{VV}(q^2)\,\frac{q^\mu q^\nu}{q^2}\rg],
\end{align}
and the bubble diagrams are defined by
\begin{align}
\label{eq:bubble_PP}
&\Pi_{PP}\lf(q^2\rg) = 6i \int \frac{d^4k}{(2\pi)^4}\ \mathrm{Tr}_D\lf[\g_5\,S(k)\,\g_5\,S(k+q)\rg], \\[0.5ex]
\label{eq:bubble_VV}
&\Pi_{VV}(q^2)\lf(g^{\mu\nu} - \frac{q^\mu q^\nu}{q^2}\rg) = \no \\
&\hs{17mm}6i \int \frac{d^4k}{(2\pi)^4}\ \mathrm{Tr}_D\lf[\g^\mu\,S(k)\,\g^\nu\,S(k+q)\rg].
\end{align}
The meson masses are given by the poles in the reduced $t$-matrices, that is
\begin{align}
\label{Eq:polepion}
1 + 2\,G_\pi\,\Pi_{PP}\bigl(q^2 = m_\pi^2\bigr) &= 0, \\  
\label{Eq:polerho}
1 + 2\,G_\rho\,\Pi_{VV}\bigl(q^2 = m_\rho^2\bigr) &= 0, \\
\label{Eq:poleomega}
1 + 2\,G_\o\,\Pi_{VV}\bigl(q^2 = m_\o^2\bigr) &= 0.
\end{align}
Expanding the full $t$-matrices about these poles gives the homogeneous Bethe-Salpeter vertices
for the $\pi$, $\rho$ and $\omega$ mesons:
\begin{align}
\G_\pi^i=\sqrt{Z_\pi}\,\g_5\,\tau_i, \quad~ \G_\rho^{\mu , i}=\sqrt{Z_\rho}\,\g^\mu\,\tau_i, \quad~ \G_\o^\mu=\sqrt{Z_\o}\,\g^\mu,
\label{Eq:bethesalpetervertices}
\end{align}
where the meson-quark-quark couplings read~\cite{Ninomiya:2014kja,Vogl:1991qt,Klevansky:1992qe} 
\begin{align}
\label{eq:Zpi}
Z_{\pi}^{-1}  &= -\,\frac{\pl}{\pl q^2}\,\Pi_{PP}(q^2)\Big\rvert_{q^2 = m_{\pi}^2}, \\[0.0ex]
\label{eq:Zrho}
Z_{\rho(\o)}^{-1} &= -\,\frac{\pl}{\pl q^2}\,\Pi_{VV}(q^2)\Big\rvert_{q^2 = m_{\rho(\o)}^2}.
\end{align}

\section{Rho Electromagnetic Form Factors \label{sec:FFs}}

\begin{figure}[tbp]
\centering\includegraphics[width=\columnwidth,clip=true,angle=0]{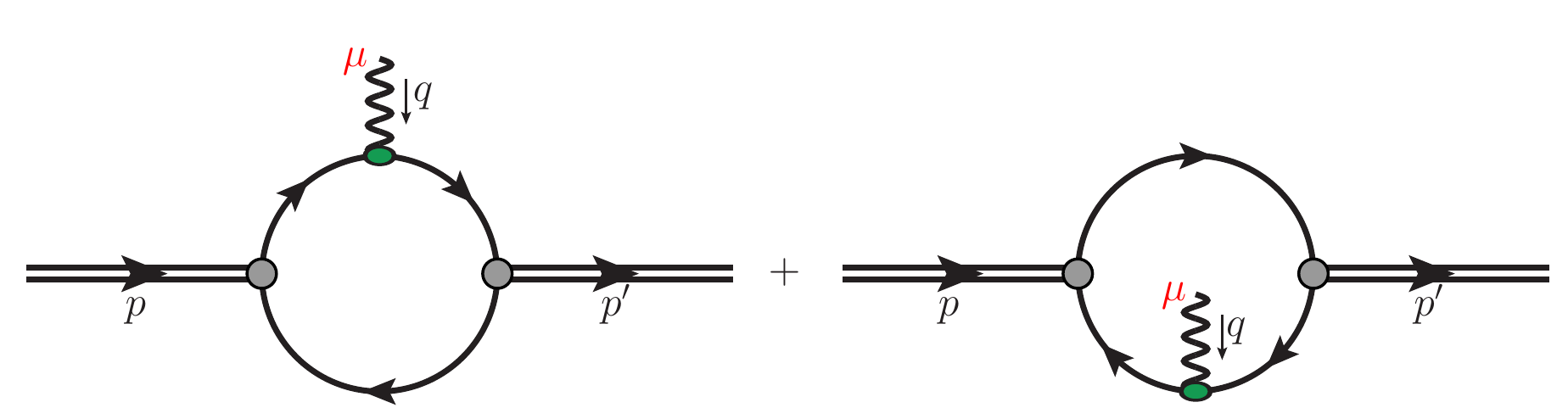}
\caption{(Colour online) Feynman diagrams representing the electromagnetic current for a meson
in our NJL model.}
\label{fig:rho_current}
\end{figure}

The electromagnetic current for a $\rho$-meson is parameterized 
by three form factors and takes the form:
\begin{align}
j^{\mu,\a\b}_{\rho}(p',p) &= \Big[g^{\a\b}F_{1\rho}(Q^2)-\frac{q^\a q^\b}{2\,m_\rho^2}\,F_{2\rho}(Q^2)\Big]\lf(p'+p\rg)^\mu \no \\
&- \lf(q^\a g^{\mu\b} - q^\b g^{\mu\a}\rg)F_{3\rho}(Q^2),
\label{eq:rhoformfactors}
\end{align}
where the polarization of the incoming and outgoing $\rho$-meson is represented by the Lorentz indices
$\a$ and $\b$, respectively, and $\mu$ is the photon polarization. From these form factors one can define 
three Sachs form factors for the $\rho$, namely, the charge [$G_C(Q^2)$]; the magnetic [$G_M(Q^2)$] and 
quadrupole [$G_Q(Q^2)$] form factors, which read
\begin{align}
G_C(Q^2) &= F_1(Q^2) + \tfrac{2}{3}\eta\, G_Q(Q^2), \label{Eq:Gc}\\
G_M(Q^2) &= F_3(Q^2), \label{Eq:GM} \\
G_Q(Q^2) &= F_1(Q^2) + (1+\eta)\,F_2(Q^2) - F_3(Q^2), \label{Eq:GQ}
\end{align}
where $\eta = \tfrac{Q^2}{4\,m_\rho^2}$ and all form factors are dimensionless. 

\begin{figure}[tbp]
\centering\includegraphics[width=\columnwidth,clip=true,angle=0]{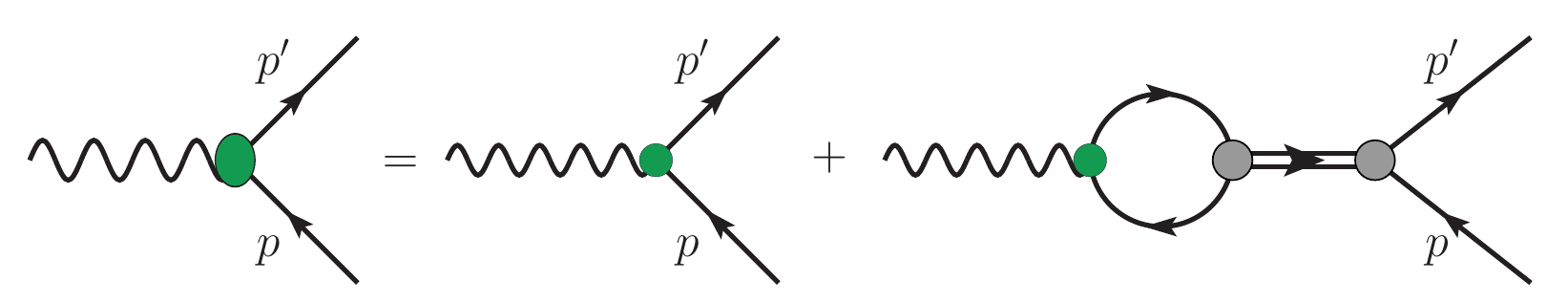}
\caption{(Colour online) Inhomogeneous BSE whose solution
gives the quark-photon vertex, represented as the large shaded oval.
The small circle depicts the pointlike 
quark-photon driving term of Eq.~\eqref{Eq:QPVertexPL}, whereas
the shaded circles with the double line represents the vector meson $t$-matrices.}
\label{fig:bse_quark}
\end{figure}

\begin{figure}[tbp]
\centering\includegraphics[width=\columnwidth,clip=true,angle=0]{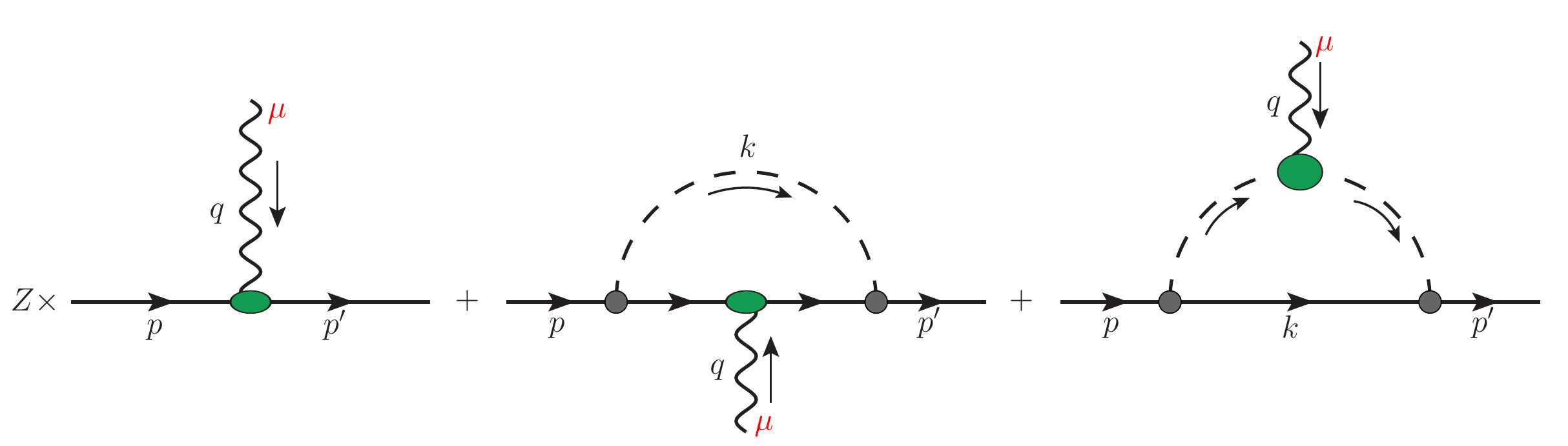}
\caption{(Colour online) Pion cloud contribution to the quark-photon vertex. The quark-photon
interaction in the first two diagrams, represented by the shaded ovals, is given by the solution 
of the inhomogeneous BSE.  The last diagram also includes the pion form factor determined
without the pion cloud on the associated dressed quarks.}
\label{fig:pion_loops_ff}
\end{figure}

In our NJL model the $\rho$ electromagnetic current is depicted in 
Fig.~\ref{fig:rho_current} and expressed by
\begin{align}
&j^{\mu,\a\b}_{\rho,ij}(p',p) = i\int\frac{d^4k}{(2\pi)^4} \no \\
&\hs{1.5mm}
\times \mathrm{Tr}\Big[\ol{\G}_\rho^{\beta,j}\,S(p'+k)\,\L^{\mu}(p',p)\,S(p+k)\,\G_\rho^{\a,i}\,S(k)\Big] \no \\
&+i\int\frac{d^4k}{(2\pi)^4} \no \\
&\hs{1.5mm}
\times \mathrm{Tr}\Big[\G_\rho^{\a,i}\,S(k-p)\,\L^{\mu}(p',p)\,S(k-p')\,\ol{\G}_\rho^{\beta,j}\,S(k)\Big],
\label{eq:hadroncurrent}
\end{align}
where the Bethe-Salpeter vertices for the $\rho$ are given in Eq.~\eqref{Eq:bethesalpetervertices},
$\L^\mu(p,p')$ is the dressed quark-photon vertex and the trace is over Dirac, color and isospin indices.
Following the calculations in Refs.~\cite{Cloet:2014rja,Ninomiya:2014kja}, we 
consider three versions of the quark-photon vertex, each of increasing sophistication;
a pointlike quark-photon vertex; a vertex given by the solution of the inhomogeneous
Bethe-Salpeter equation (illustrated in Fig.~\ref{fig:bse_quark}); and finally a quark-photon 
vertex which includes the pion cloud at the quark level (see Fig.~\ref{fig:pion_loops_ff}).

The pointlike quark-photon is simply given by
\begin{align}
\L^{(\text{PL})\mu}(p,p') = \lf[\frac{1}{6} + \frac{\tau_3}{2}\rg]\g^{\mu},
\label{Eq:QPVertexPL}
\end{align}
where $\tfrac{1}{6} + \tfrac{\tau_3}{2}$ is the quark charge operator.
Projecting onto flavour sectors the vertex is separated into two components:
\begin{align}
\L^{(\text{PL})\mu}(p,p') = \lf[e_u\,\frac{1+\tau_3}{2} + e_d\,\frac{1-\tau_3}{2}\rg]\g^\mu,
\label{Eq:QgPVertexPLFS}
\end{align}
where $e_u$ and $e_d$ are the charges of the $u$ and $d$ quarks, respectively.

In general the quark-photon vertex is dressed by $\bar{q}q$ interactions in the
vector channel and in the NJL model this dressing is described by the corresponding inhomogeneous 
Bethe-Salpeter equation (see Fig.~\ref{fig:bse_quark}).
From the NJL Lagrangian of Eq.~\eqref{eq:lagrangian} the contributions to this
vertex come from the neutral vector mesons ($\rho^0$ and $\omega$). In the on-shell 
approximation for the external quarks, the solution of the inhomogeneous
Bethe-Salpeter equation of Fig.~\ref{fig:bse_quark} is
\begin{align}
\L^{(\text{bse})\mu}(p,p') = 
\lf[\frac{1}{6}\, F_{1\omega}(q^2) + \frac{\tau_3}{2}\, F_{1\rho}(q^2)\rg]\g^{\mu},
\label{Eq:QPVertexVMD}
\end{align}
where the dressed quark form factors are
\begin{align}
F_{1\omega(\rho)}(q^2) = \frac{1}{1 + 2\,G_{\omega(\rho)}\,\Pi_{VV}(q^2)}.
\label{Eq:vecmFF}
\end{align}
Note, with the Lagrangian of Eq.~\eqref{eq:lagrangian} the inhomogeneous Bethe-Salpeter equation
does not generate a Pauli form factor for the dressed quarks.
Again projecting into flavour sectors gives
\begin{align}
\L^{(\text{bse})\mu}(p,p') = \lf[F^{\text{bse}}_{1U}(q^2)\,\frac{1+\tau_3}{2} + F^{\text{bse}}_{1D}(q^2)\,\frac{1-\tau_3}{2}\rg]\g^\mu,
\label{Eq:QPVertexVMDFS}
\end{align}
where the dressed quark form factors are given by~\cite{Cloet:2014rja}
\begin{align}
\label{eq:f1u}
F^{\text{bse}}_{1U}(Q^2) &= \frac{1}{6}F_{1\omega}(Q^2) + \frac{1}{2}F_{1\rho}(Q^2), \\
\label{eq:f1d}
F^{\text{bse}}_{1D}(Q^2) &= \frac{1}{6}F_{1\omega}(Q^2) - \frac{1}{2}F_{1\rho}(Q^2).
\end{align}

Finally we include pion loop corrections to the quark-photon vertex, as illustrated in
Fig.~\ref{fig:pion_loops_ff}, which give a vertex of the form
\begin{align}
\L^{\mu}(p,p') = \L^{\mu}_U(p,p')\,\frac{1+\tau_3}{2} + \L^{\mu}_D(p,p')\,\frac{1-\tau_3}{2},
\label{eq:vertexpion}
\end{align}
where the flavour sector vertices ($Q=U,\,D$) read
\begin{align}
\L^{\mu}_Q(p,p') = \g^{\mu}\, F_{1Q}(q^2) + \frac{i\s^{\mu\mu}q_{\nu}}{2\,M}\,F_{2Q}(Q^2).
\label{eq:pivertex}
\end{align}
Note that the pion cloud generates a Pauli form factor for the dressed quarks and 
that in obtaining Eq.~\eqref{eq:pivertex} we have assumed the external quark lines are on-shell.
The dressed quark form factors now read~\cite{Cloet:2014rja}
\begin{align}
\label{eq:F_1U}
F_{1U} &= Z\lf[\tfrac{1}{6}F_{1\omega} + \tfrac{1}{2}F_{1\rho}\rg]+\lf[F_{1\omega} - F_{1\rho}\rg]f_1^{q}+F_{1\rho}f_{1}^{\pi},\\
\label{eq:F_1D}
F_{1D} &= Z\lf[\tfrac{1}{6}F_{1\omega} - \tfrac{1}{2}F_{1\rho}\rg]+\lf[F_{1\omega} + F_{1\rho}\rg]f_1^{q}-F_{1\rho}f_{1}^{\pi},\\
\label{eq:F_2U}
F_{2U} &= \lf[F_{1\omega} - F_{1\rho}\rg]f_2^{q}+F_{1\rho}f_{2}^{\pi},\\
\label{eq:F_2D}
F_{2D} &= \lf[F_{1\omega} + F_{1\rho}\rg]f_2^{q}-F_{1\rho}f_{2}^{\pi},
\end{align}
where for clarity we have dropped the explicit $Q^2$ dependence. The
renormalization factor $Z$ is given by
\begin{align}
Z = 1 + \lf.\frac{\pl\S(p)}{\pl\sh{p}}\rg|_{\sh{p}=M},
\label{eq:pion_correction}
\end{align}
where $\S(p)$ is the self-energy from the pion cloud on a dressed quark:
\begin{align}
\S(p)=-\int\frac{d^4k}{(2\pi)^4}\,\g_5\,\tau_i\,S(p-k)\,\g_5\,\tau_i\,\tau_{\pi}(k).
\label{eq:pion_self_energy}
\end{align}
Here the pion propagator is approximated by its pole form
%
\begin{align}
\tau_\pi(k) \to \frac{i\,Z_\pi}{p^2 - m_{\pi}^2+i\epsilon}.
\end{align}
The contributions of the pion cloud to the quark-photon vertex are contained
in the functions $f_i^{q}(Q^2)$ and $f_i^{\pi}(Q^2)~(i=1,\,2)$ of Eqs.~\eqref{eq:F_1U}-\eqref{eq:F_2D}.
These body form factors are
associated with the second and third diagrams in Fig.~\ref{fig:pion_loops_ff}, 
which are respectively expressed as
\begin{align}
\label{eq:f12q}
\L_Q^{(q)\mu}(p',p)  &= \g^{\mu}\,f_1^{q}(q^2) + \frac{i\s^{\mu\nu}q_{\nu}}{2\,M}\,f_2^{q}(q^2), \\
\label{e1:f12pi}
\L_Q^{(\pi)\mu}(p',p) &= \g^{\mu}\,f_1^{\pi}(q^2) + \frac{i\s^{\mu\nu}q_{\nu}}{2\,M}\,f_2^{\pi}(q^2).
\end{align}
The analytic expressions read
\begin{align}
\label{eq:L1q}
\L_Q^{(q)\mu}(p',p) &=\!\int\!\!\frac{d^4k}{(2\pi)^4}\,\g_5\,iS(p'-k)\g^{\mu}\,iS(p-k)\g_5\,\tau_{\pi}(k), \allowdisplaybreaks\\
\label{eq:L1pi}
\L_Q^{(\pi)\mu}(p',p) &= Z_\pi^{-1}\,(p'+p)^{\mu}\,F_{\pi}^{(\text{PL})}(q^2)\no\\
&\hs{-7mm}
\times\int\frac{d^4k}{(2\pi)^4}\ \g_5\,\tau_\pi(p'-k)\,\tau_{\pi}(p-k)\,\g_5\,iS(k),
\end{align}
where $F_{\pi}^{(\text{PL})}(q^2)$ is the pion form factor determined with a pointlike
quark-photon vertex. 

For the full calculation of the $\rho$-meson form factors we use 
Eq.~\eqref{eq:hadroncurrent} and the quark-photon vertex given in Eq.~\eqref{eq:vertexpion}. 
For the $\rho^+$ form factors this gives
\begin{align}
F_{i\rho^+}(Q^2) &= \lf[F_{1U}(Q^2) - F_{1D}(Q^2)\rg]f_{i}^{V}(Q^2)\no\\
&+\lf[F_{2U}(Q^2) - F_{2D}(Q^2)\rg]f_{i}^{T}(Q^2),
\label{eq:dressedrhoff}
\end{align}
where $i = 1,2,3$ indicates each of the three form factors of Eq.~\eqref{eq:rhoformfactors}.
The body form factors $f_{i}^{V}$ are associated with the vector part ($\g^{\mu}$) of the 
quark-photon vertex in Eq.~\eqref{eq:pivertex}, while $f_{i}^{T}$ are the 
body form factors associated with the tensor coupling ($\frac{i\s^{\mu\nu}q_{\nu}}{2\,M}$) in
Eq.~\eqref{eq:pivertex}. To obtain the $\rho$-meson form factors
that result only from the inhomogeneous BSE quark-photon vertex we then simply set $Z=1$ and the pion cloud contributions ($f_1^{q}(Q^2)$, $f_{1}^{\pi}(Q^2)$, etc) to zero. Finally, the $\rho$ form factors for a pointlike
quark-photon vertex are then obtained by setting $F_{1\omega} = F_{1\rho} = 1$. Note, all
loop integrals are regularized using the proper-time scheme, with both an infrared 
and ultraviolet cutoff, except those of Eqs.~\eqref{eq:pion_self_energy}, \eqref{eq:L1q} 
and \eqref{eq:L1pi}, where we take the infrared cutoff ($\L_{IR}$) to zero as the pion 
should not be confined.

\section{Results\label{sec:results}}

\begin{table}[tbp]
\addtolength{\tabcolsep}{2.5pt}
\addtolength{\extrarowheight}{2.2pt}
\begin{tabular}{ccccccccc}
\hline\hline
$M$ & $\L_{IR}$ & $\L_{UV}$ & $G_{\pi}$ & $G_{\rho}$ & $G_{\omega}$ & $Z_{\pi}$ & $Z_{\rho}$ & $Z_{\omega}$ \\[0.2em]
\hline
0.4 &   0.24   &   0.645  &   19.04  &   11.04  &   10.41    &   17.85  &    6.96   &    6.63    \\
\hline\hline
\end{tabular}
\caption{Parameters of the model together with the effective couplings computed from
Eqs.~\eqref{eq:Zpi}-\eqref{eq:Zrho}. The masses are in units of GeV, the Lagrangian couplings
in units of GeV$^{-2}$ and the effective couplings are dimensionless.} 
\label{tab:parameters}
\end{table}

The parameters of our model are the dressed quark mass $M$; the regularization
cutoffs $\L_{UV}$ and $\L_{IR}$; and the Lagrangian couplings $G_{\pi}$, $G_{\rho}$ and $G_{\omega}$.
For consistency with previous work
we set $M=0.4$ GeV (in the physical limit: $m_\pi = 140\,$MeV)
and $\L_{IR} = 0.24\,$GeV~\cite{Cloet:2014rja,Ninomiya:2014kja,Carrillo-Serrano:2014zta}. 
The ultraviolet cutoff $\L_{UV}$ is fit to 
the physical value of the pion decay constant and the couplings $G_{\pi}$, $G_{\rho}$, and 
$G_{\omega}$ are fit to the physical masses of the $\pi$, $\rho$ and $\omega$ mesons using Eqs.~\eqref{Eq:polepion}-\eqref{Eq:poleomega}. 
The values of these parameters, together with the quark-meson couplings of Eqs.~\eqref{eq:Zpi}-\eqref{eq:Zrho}, 
are given in Tab.~\ref{tab:parameters}.

Our purpose here is to compare results within this NJL model with other calculations, for example, constituent quark models~\cite{Gudino:2013jaa,Chung:1988my,Cardarelli:1994yq,deMelo:1997hh,
Choi:1997iq,Melikhov:2001pm,Jaus:2002sv,Choi:2004ww,Biernat:2014dea}, 
QCD sum rules~\cite{Aliev:2003ba,Samsonov:2003hs,Braguta:2004kx,Aliev:2004uj},
Dyson-Schwinger equations~\cite{Hawes:1998bz,Bhagwat:2006pu,Roberts:2011wy,Pitschmann:2012by}
and the recent lattice QCD studies~\cite{Owen:2015gva,Shultz:2015pfa}. 
We first focus on static electromagnetic quantities for the $\rho^+$ meson and consider
the magnetic moment ($\mu_\rho$), quadrupole moment ($\calQ_\rho$) and rms charge radius ($\langle r_C^2 \rangle$).
These observables are defined by the Sachs form factors given in Eqs.~\eqref{Eq:Gc}-\eqref{Eq:GQ}, 
where the magnetic moment in nuclear magnetons ($\mu_N$) is given by $\mu_\rho = G_{M}(0)\tfrac{M_N}{m_\rho}$,
with $M_N$ the physical nucleon mass and (for comparison with lattice data) $m_{\rho}$ 
is the $\rho$ mass evaluated at a particular pion mass; the quadrupole moment 
in units of $e/m^2_{\rho}$ is given by $\calQ_\rho = G_{Q}(0)$; and finally the charge radius is defined by
\begin{align}
\lf< r_C^2\rg> = \lf.-6\,\frac{\pl\, G_{C}(Q^2)}{\pl Q^2}\rg|_{Q^2=0}.
\label{eq:radius}
\end{align}
In Tab.~\ref{tab:magmom_quadmom_comparison} we summarize results for the
$\langle r_C^2\rangle$, $\mu_\rho$ and $\calQ_\rho$ of the $\rho^+$ from various theoretical 
approaches, together with our calculations using the most sophisticated 
quark-photon vertex of Eq.~\eqref{eq:vertexpion} (BSE + pion cloud).
In general including the dressing of the quark-photon vertex by the 
BSE and the pion cloud increases the magnitude
of $\mu_\rho$ by 24\%, $\calQ_\rho$ by 22\% and $\langle r_{C}^2 \rangle$ by 16\%~\cite{Cloet:2014rja}.

\begin{table}[tbp]
\addtolength{\tabcolsep}{4.5pt}
\addtolength{\extrarowheight}{2.2pt}
\begin{tabular}{lcccc}
\hline\hline
Reference & $\langle r_C^{2} \rangle$(fm$^2$) & $\mu_\rho$ ($\mu_{N}$)  &
$\calQ_\rho$ (fm$^2$) \\ 
\hline
This work                            & {\color{red} 0.67} &  {\color{red}
3.14}   &  {\color{red} -0.070} \\
Garcia Gudi\~no~\cite{Gudino:2013jaa} &  --        &  2.6(6)     &  --    \\
Cardarelli~\cite{Cardarelli:1994yq}   & 0.35       &  2.76       &  -0.024 \\
De Melo~\cite{deMelo:1997hh}          & 0.37       &  2.61       &  -0.052 \\
Melikhov~\cite{Melikhov:2001pm}       & 0.33       &  2.87       &  -0.031 \\
Jaus~\cite{Jaus:2002sv}               &  --        &  2.23       &  -0.022 \\
Choi~\cite{Choi:2004ww}               &  --        &  2.34       &  -0.028 \\
Biernat~\cite{Biernat:2014dea}        &  --        &  2.68       &  -0.027 \\
Samsonov~\cite{Samsonov:2003hs}       &  --        &  2.4(4)     &  --    \\
Aliev~\cite{Aliev:2004uj}             &  --        &  2.8(6)     &    --     \\
Hawes~\cite{Hawes:1998bz}             & 0.37       &  3.28       &  -0.055 \\
Bhagwat~\cite{Bhagwat:2006pu}         & 0.54       &  2.54       &  -0.026 \\
Roberts~\cite{Roberts:2011wy}         & 0.31       &  2.14       &  -0.037 \\
Pitschmann~\cite{Pitschmann:2012by}   &  --        &  2.13       &    --     \\
Owen~\cite{Owen:2015gva}              & 0.670(68)  &  2.613(97)  &  -0.0452(61) \\
Shultz~\cite{Shultz:2015pfa}          & 0.30(6)    &  2.00(9)    &  -0.020(4)
\\
\hline\hline
\end{tabular}
\caption{Comparison of the $\rho^+$ charge radius, magnetic moment and
quadrupole moment for various
theoretical approaches: phenomenological models~\cite{Gudino:2013jaa}, constituent quark
models~\cite{Cardarelli:1994yq,deMelo:1997hh,Melikhov:2001pm,Jaus:2002sv,Choi:2004ww,Biernat:2014dea},
QCD sum rules~\cite{Samsonov:2003hs,Aliev:2004uj},
Dyson Schwinger
equations~\cite{Hawes:1998bz,Bhagwat:2006pu,Roberts:2011wy,Pitschmann:2012by}
and lattice QCD~\cite{Owen:2015gva,Shultz:2015pfa}.
The lightest pion mass used in the lattice calculation in
Ref.~\cite{Owen:2015gva} is $m_\pi^2=0.026$ GeV$^2$, whereas for
Ref.~\cite{Shultz:2015pfa} it is $m_\pi^2=0.49$ GeV$^2$.}
\label{tab:magmom_quadmom_comparison}
\end{table}

\begin{figure}[tbp]
  \centering
  \resizebox{0.5\textwidth}{!}{\setlength{\unitlength}{0.0500bp}%
  \begin{picture}(7200.00,5040.00)%
      \put(990,704){\makebox(0,0)[r]{\strut{}\large 0.7}}%
      \put(990,1383){\makebox(0,0)[r]{\strut{}\large 0.75}}%
      \put(990,2061){\makebox(0,0)[r]{\strut{}\large 0.8}}%
      \put(990,2740){\makebox(0,0)[r]{\strut{}\large 0.85}}%
      \put(990,3418){\makebox(0,0)[r]{\strut{}\large 0.9}}%
      \put(990,4097){\makebox(0,0)[r]{\strut{}\large 0.95}}%
      \put(990,4775){\makebox(0,0)[r]{\strut{}\large 1}}%
      \put(1122,484){\makebox(0,0){\strut{}\large 0}}%
      \put(1832,484){\makebox(0,0){\strut{}\large 0.05}}%
      \put(2542,484){\makebox(0,0){\strut{}\large 0.1}}%
      \put(3252,484){\makebox(0,0){\strut{}\large 0.15}}%
      \put(3963,484){\makebox(0,0){\strut{}\large 0.2}}%
      \put(4673,484){\makebox(0,0){\strut{}\large 0.25}}%
      \put(5383,484){\makebox(0,0){\strut{}\large 0.3}}%
      \put(6093,484){\makebox(0,0){\strut{}\large 0.35}}%
      \put(6803,484){\makebox(0,0){\strut{}\large 0.4}}%
      \put(286,2739){\rotatebox{-270}{\makebox(0,0){\Large $m_{\rho}$ (GeV)}}}%
      \put(3962,154){\makebox(0,0){\Large $m_{\pi}^2$ (GeV$^2$)}}%
      \put(5099,1896){\makebox(0,0)[r]{\strut{}NJL}}%
      \put(5099,1566){\makebox(0,0)[r]{\strut{}lattice QCD}}%
    \put(0,0){\includegraphics{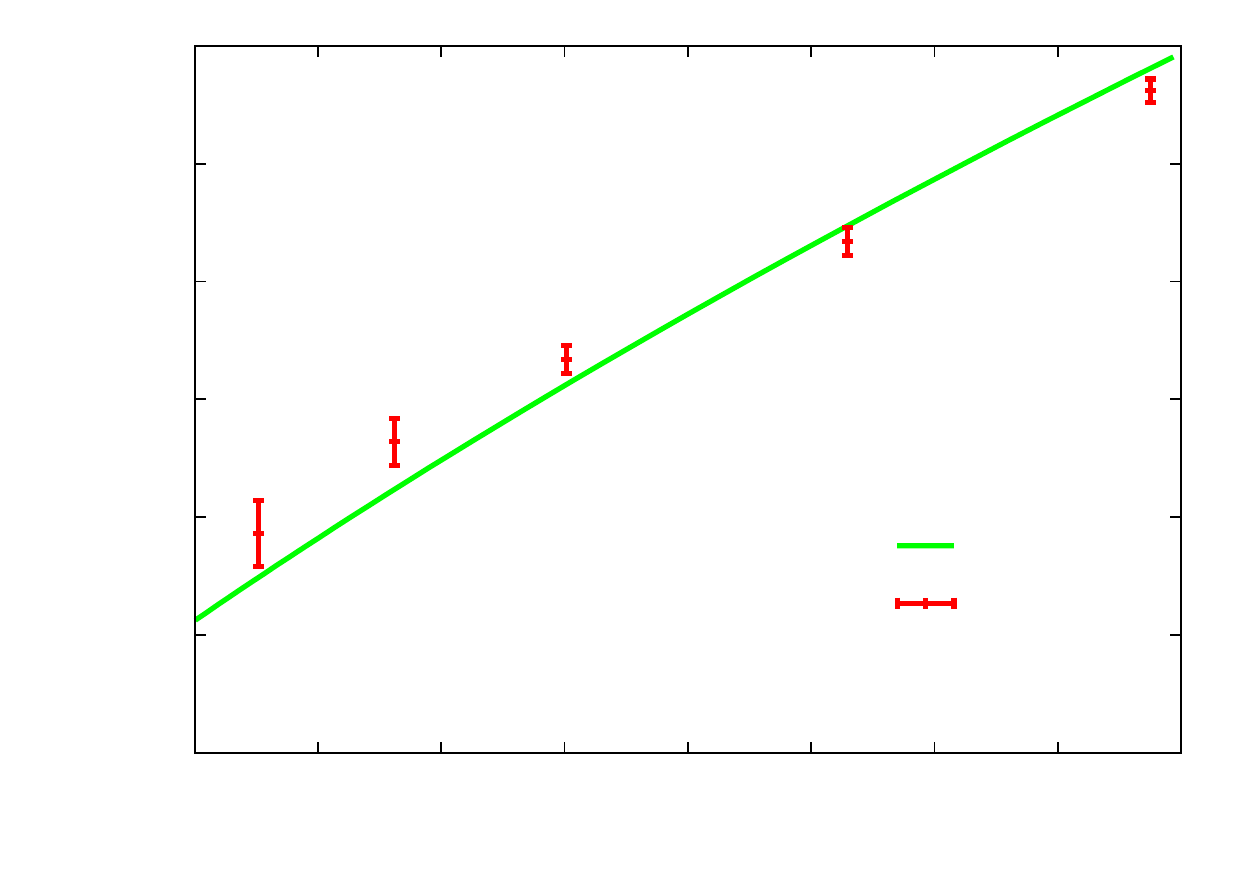}}%
  \end{picture}%
}
  \caption{(Colour online) NJL model results for the $\rho$-meson mass versus $m_{\pi}^2$.
Comparison is made with lattice results from Ref.~\cite{Owen:2015gva}. }
  \label{fig:MrVsMpi2}
\end{figure}
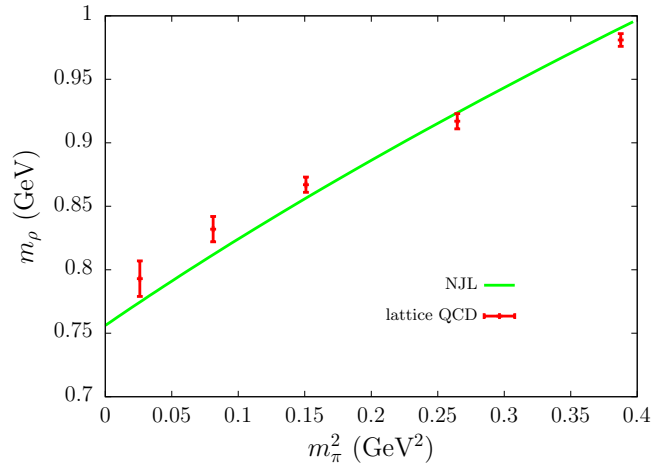

In comparing our results with lattice QCD we focus on the lattice simulation from
Ref.~\cite{Owen:2015gva}, as they extend to the lightest pion mass, namely, $m_\pi = 161\,$MeV.
Our computations as functions of $m_\pi^2$ are performed by keeping the regularization parameters
($\L_{IR}$ and $\L_{UV}$) and the couplings ($G_\pi$, $G_\rho$ and $G_\o$) fixed, and varying
the current quark mass that enters the gap equation. Results for the $\rho$ mass 
as a function of $m_\pi^2$ (or equivalently the current quark mass) are presented in Fig.~\ref{fig:MrVsMpi2},
where we find remarkable agreement between our NJL calculation and the lattice results
of Ref.~\cite{Owen:2015gva}.

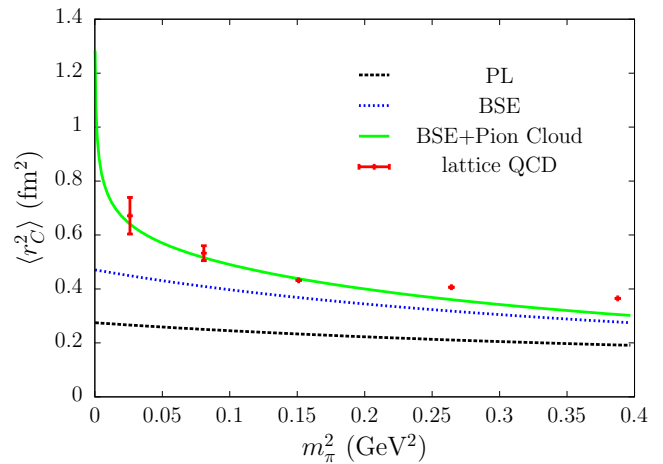
\begin{figure}[bp]
  \centering
  \resizebox{0.5\textwidth}{!}{  \setlength{\unitlength}{0.0500bp}%
  \begin{picture}(7200.00,5040.00)%
      \put(858,704){\makebox(0,0)[r]{\strut{}\large 0}}%
      \put(858,1286){\makebox(0,0)[r]{\strut{}\large 0.2}}%
      \put(858,1867){\makebox(0,0)[r]{\strut{}\large 0.4}}%
      \put(858,2449){\makebox(0,0)[r]{\strut{}\large 0.6}}%
      \put(858,3030){\makebox(0,0)[r]{\strut{}\large 0.8}}%
      \put(858,3612){\makebox(0,0)[r]{\strut{}\large 1}}%
      \put(858,4193){\makebox(0,0)[r]{\strut{}\large 1.2}}%
      \put(858,4775){\makebox(0,0)[r]{\strut{}\large 1.4}}%
      \put(990,484){\makebox(0,0){\strut{}\large 0}}%
      \put(1717,484){\makebox(0,0){\strut{}\large 0.05}}%
      \put(2443,484){\makebox(0,0){\strut{}\large 0.1}}%
      \put(3170,484){\makebox(0,0){\strut{}\large 0.15}}%
      \put(3897,484){\makebox(0,0){\strut{}\large 0.2}}%
      \put(4623,484){\makebox(0,0){\strut{}\large 0.25}}%
      \put(5350,484){\makebox(0,0){\strut{}\large 0.3}}%
      \put(6076,484){\makebox(0,0){\strut{}\large 0.35}}%
      \put(6803,484){\makebox(0,0){\strut{}\large 0.4}}%
      \put(286,2739){\rotatebox{-270}{\makebox(0,0){\Large $\langle r_{C}^2\rangle$ (fm$^2$)}}}%
      \put(3896,154){\makebox(0,0){\Large $m_{\pi}^2$ (GeV$^2$)}}%
      \put(5350,4174){\makebox(0,0){\large PL}}%
      \put(5350,3844){\makebox(0,0){\large BSE}}%
      \put(5350,3514){\makebox(0,0){\large BSE+Pion Cloud}}%
      \put(5350,3184){\makebox(0,0){\large lattice QCD}}%
    \put(0,0){\includegraphics{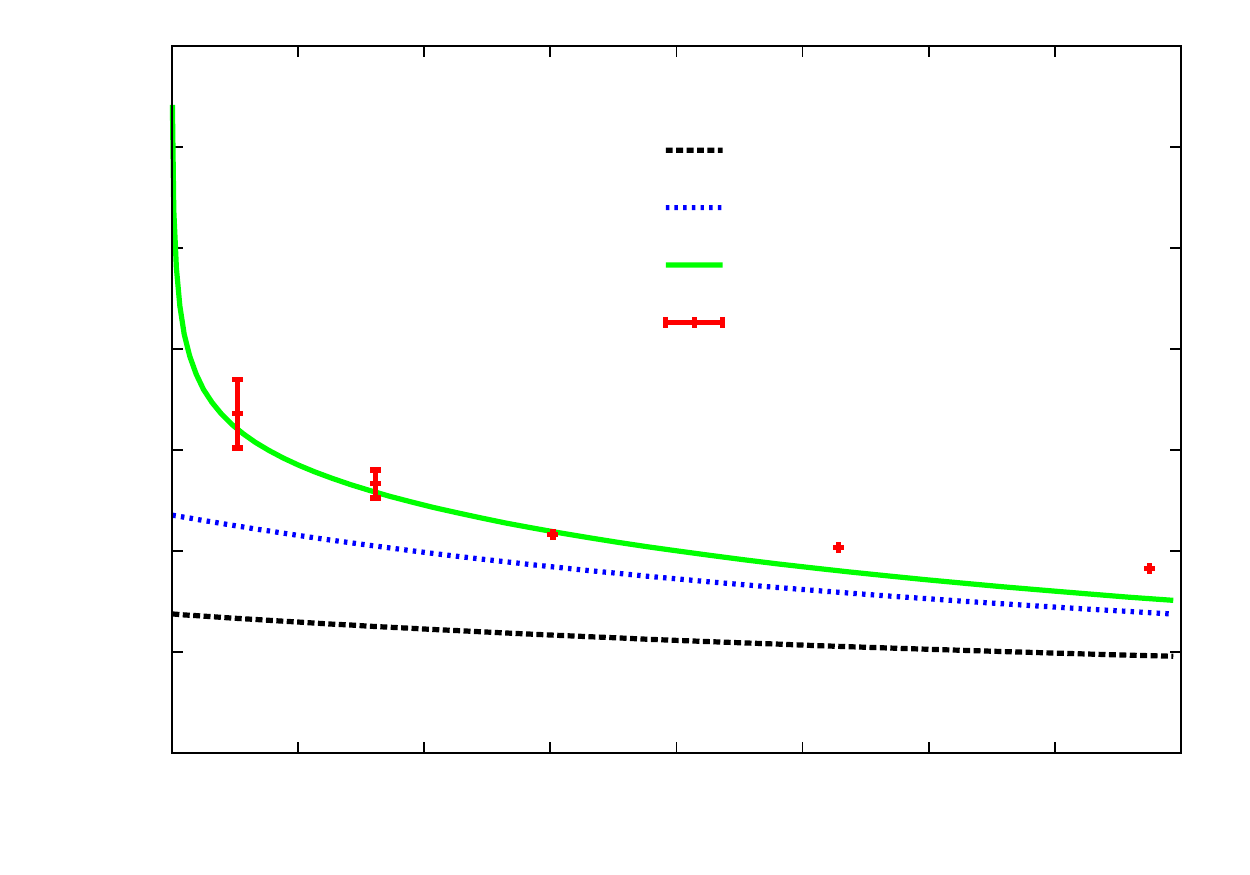}}%
  \end{picture}%
}
\caption{(Colour online) The squared charge radius $\langle r_C^2 \rangle$ for the $\rho^+$ meson
computed using the three levels of sophistication for the quark-photon vertex: pointlike (PL),
using the inhomogeneous BSE (BSE) and including the pion cloud (BSE + pion cloud).
Comparison is made with lattice results from Ref.~\cite{Owen:2015gva}. }
  \label{fig:radius}
\end{figure}

At the physical pion mass our values for $\langle r_C^2 \rangle$ (see Tab.~\ref{tab:magmom_quadmom_comparison})
differ significantly from the constituent quark models, one of the Dyson-Schwinger calculations and the result quoted
in the lattice QCD computation of Ref.~\cite{Shultz:2015pfa}. 
Better agreement is seen with the Dyson-Schwinger equation calculation of Ref.~\cite{Bhagwat:2006pu}. 
Our result for $\langle r_C^2\rangle$ is however very similar to the lattice QCD
value obtained in Ref.~\cite{Owen:2015gva} for a pion mass of around $161$ MeV.  We see
that in Fig.~\ref{fig:radius} their $\langle r_C^2 \rangle$ lies around $0.67$ fm$^2$, possibly
reaching $0.7$ fm$^2$ in the physical limit.
On the other hand the lattice QCD simulation of Ref.~\cite{Shultz:2015pfa} uses a very large pion mass of $700$ MeV, which
explains its lower value for $\langle r_C^2\rangle$, evident from the $m_\pi^2$ dependence of the 
lattice points in Fig.~\ref{fig:radius}. The dependence
of $\langle r_C^2 \rangle$ on $m_\pi^2$ in our NJL calculation, once the inhomogeneous BSE 
and pion cloud contributions have been included, shows remarkable agreement with the lattice results
of Ref.~\cite{Owen:2015gva}. One sees that the pion cloud contributions have become negligible for
$m_\pi^2 \gtrsim 0.4\,$GeV$^2$.

\begin{figure}[tbp]
  \centering\resizebox{0.5\textwidth}{!}{\setlength{\unitlength}{0.0500bp}%
  \begin{picture}(7200.00,5040.00)%
      \put(858,1074){\makebox(0,0)[r]{\strut{}\large 2}}%
      \put(858,1999){\makebox(0,0)[r]{\strut{}\large 2.5}}%
      \put(858,2925){\makebox(0,0)[r]{\strut{}\large 3}}%
      \put(858,3850){\makebox(0,0)[r]{\strut{}\large 3.5}}%
      \put(858,4775){\makebox(0,0)[r]{\strut{}\large 4}}%
      \put(990,484){\makebox(0,0){\strut{}\large 0}}%
      \put(1717,484){\makebox(0,0){\strut{}\large 0.05}}%
      \put(2443,484){\makebox(0,0){\strut{}\large 0.1}}%
      \put(3170,484){\makebox(0,0){\strut{}\large 0.15}}%
      \put(3897,484){\makebox(0,0){\strut{}\large 0.2}}%
      \put(4623,484){\makebox(0,0){\strut{}\large 0.25}}%
      \put(5350,484){\makebox(0,0){\strut{}\large 0.3}}%
      \put(6076,484){\makebox(0,0){\strut{}\large 0.35}}%
      \put(6803,484){\makebox(0,0){\strut{}\large 0.4}}%
      \put(286,2739){\rotatebox{-270}{\makebox(0,0){\Large $\mu_{\rho}$ $(\mu_N)$}}}%
      \put(3896,154){\makebox(0,0){\Large $m_{\pi}^2$ (GeV$^2$)}}%
    \put(0,0){\includegraphics{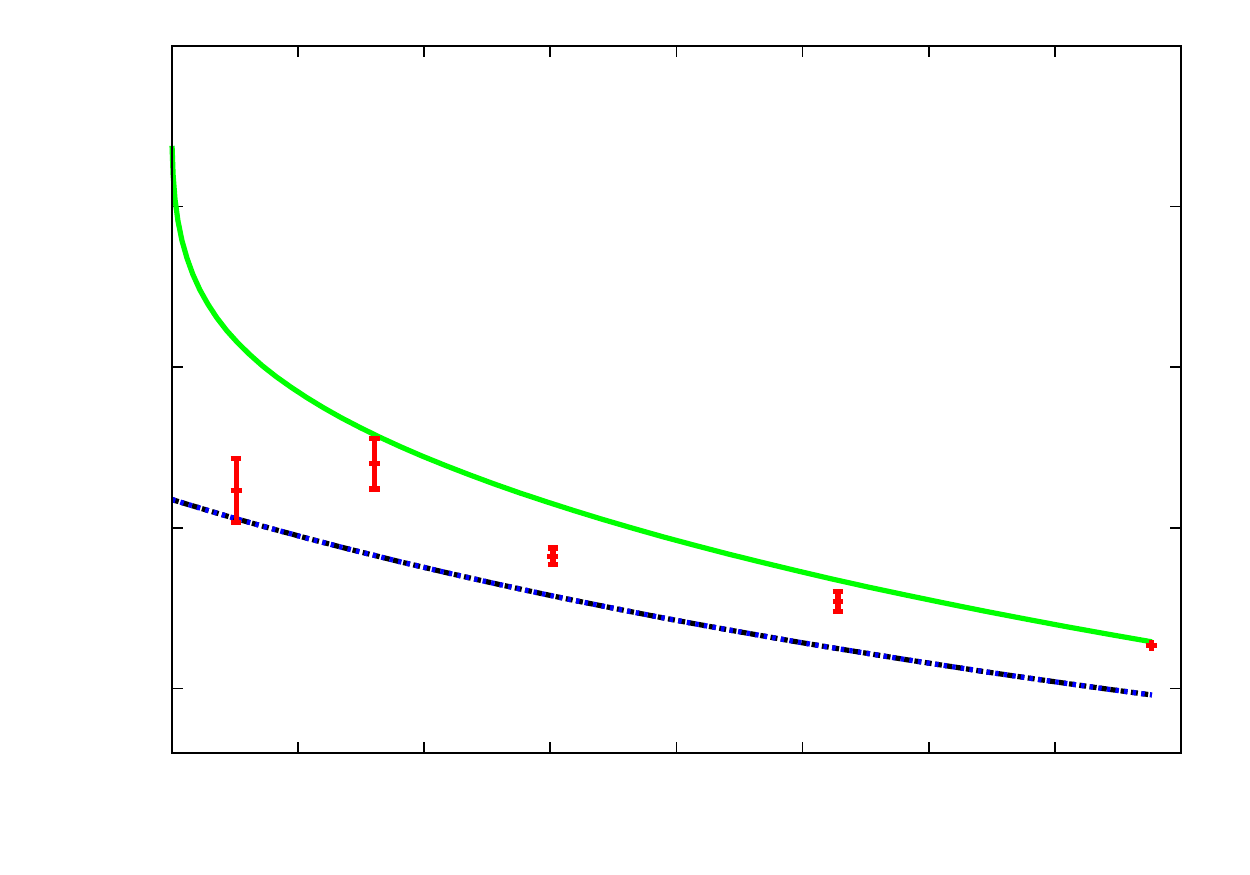}}%
  \end{picture}%
}
\caption{(Colour online) The $\rho^+$ magnetic moment ($\mu_{\rho}$) versus $m_{\pi}^2$. 
The curves have the same conventions as Fig.~\ref{fig:radius}.}
\label{fig:mu}
\vspace{0.8em}
\centering\resizebox{0.5\textwidth}{!}{\setlength{\unitlength}{0.0500bp}%
  \begin{picture}(7200.00,5040.00)%
      \put(1122,871){\makebox(0,0)[r]{\strut{}\large -0.08}}%
      \put(1122,1429){\makebox(0,0)[r]{\strut{}\large -0.07}}%
      \put(1122,1987){\makebox(0,0)[r]{\strut{}\large -0.06}}%
      \put(1122,2544){\makebox(0,0)[r]{\strut{}\large -0.05}}%
      \put(1122,3102){\makebox(0,0)[r]{\strut{}\large -0.04}}%
      \put(1122,3660){\makebox(0,0)[r]{\strut{}\large -0.03}}%
      \put(1122,4217){\makebox(0,0)[r]{\strut{}\large -0.02}}%
      \put(1122,4775){\makebox(0,0)[r]{\strut{}\large -0.01}}%
      \put(1254,484){\makebox(0,0){\strut{}\large 0}}%
      \put(1948,484){\makebox(0,0){\strut{}\large 0.05}}%
      \put(2641,484){\makebox(0,0){\strut{}\large 0.1}}%
      \put(3335,484){\makebox(0,0){\strut{}\large 0.15}}%
      \put(4029,484){\makebox(0,0){\strut{}\large 0.2}}%
      \put(4722,484){\makebox(0,0){\strut{}\large 0.25}}%
      \put(5416,484){\makebox(0,0){\strut{}\large 0.3}}%
      \put(6109,484){\makebox(0,0){\strut{}\large 0.35}}%
      \put(6803,484){\makebox(0,0){\strut{}\large 0.4}}%
      \put(286,2739){\rotatebox{-270}{\makebox(0,0){\Large $\calQ_{\rho}$ (fm$^2$)}}}%
      \put(4028,154){\makebox(0,0){\Large $m_{\pi}^2$ (GeV$^2$)}}%
      \put(2225,66677){\makebox(0,0){\large $G_{C}$}}%
      \put(2225,133597){\makebox(0,0){\large $G_{M}$}}%
      \put(2225,-28126){\makebox(0,0){\large $G_{Q}$}}%
    \put(0,0){\includegraphics{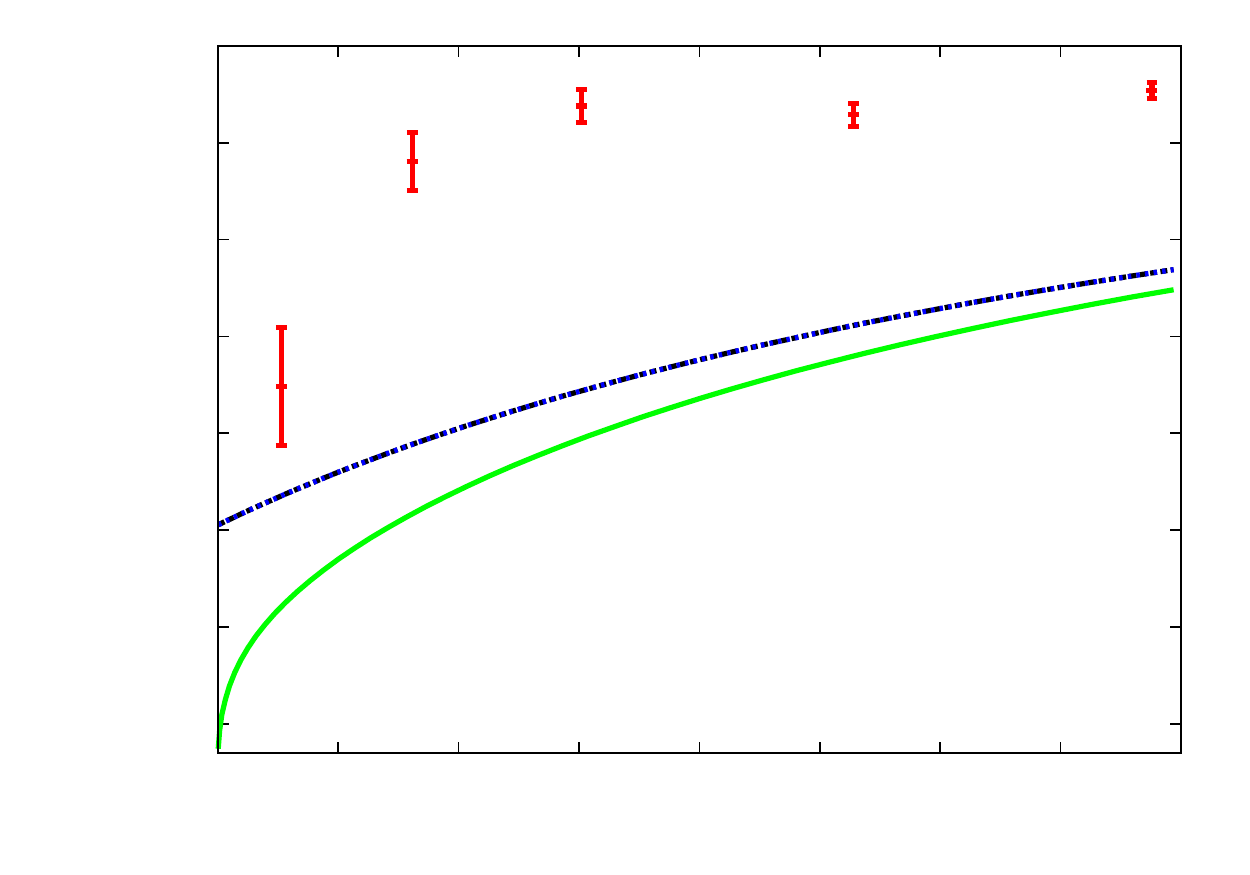}}%
  \end{picture}%
}
\caption{(Colour online) The $\rho^+$ quadrupole moment ($\calQ_{\rho}$)
versus $m_{\pi}^2$. The curves have the same conventions as Fig.~\ref{fig:radius}.}
\label{fig:Q}
\end{figure}

For the $\rho^+$ magnetic moment ($\mu_\rho$) the values obtained by the
constituent quark models are consistently smaller than our result of $\mu_\rho = 3.14\,\mu_N$, 
the closest being $\mu_\rho = 2.87\,\mu_N$ from Ref.~\cite{Melikhov:2001pm}. 
The earlier Dyson-Schwinger equation study in Ref.~\cite{Hawes:1998bz} shows good agreement with our work.
For the lattice simulation of Ref.~\cite{Owen:2015gva} the discrepancy with our result is sizeable near the physical limit.
However, the evolution of our result with $m_{\pi}^2$ shown in Fig.~\ref{fig:mu} is in good agreement with the lattice QCD calculations except at their lightest pion mass. Again, as in the case of $\langle r_C^2 \rangle$, the effect of the 
large $m_{\pi}$ in Ref.~\cite{Shultz:2015pfa} is to produce a small value of $\mu_\rho$, 
as evident from Fig.~\ref{fig:mu}.

\begin{figure}[tbp]
\centering\resizebox{0.50\textwidth}{!}{\setlength{\unitlength}{0.0500bp}%
  \begin{picture}(7200.00,5040.00)%
      \put(990,704){\makebox(0,0)[r]{\strut{}\large 0.6}}%
      \put(990,1383){\makebox(0,0)[r]{\strut{}\large 0.65}}%
      \put(990,2061){\makebox(0,0)[r]{\strut{}\large 0.7}}%
      \put(990,2740){\makebox(0,0)[r]{\strut{}\large 0.75}}%
      \put(990,3418){\makebox(0,0)[r]{\strut{}\large 0.8}}%
      \put(990,4097){\makebox(0,0)[r]{\strut{}\large 0.85}}%
      \put(990,4775){\makebox(0,0)[r]{\strut{}\large 0.9}}%
      \put(1122,484){\makebox(0,0){\strut{}\large 0}}%
      \put(1832,484){\makebox(0,0){\strut{}\large 0.05}}%
      \put(2542,484){\makebox(0,0){\strut{}\large 0.1}}%
      \put(3252,484){\makebox(0,0){\strut{}\large 0.15}}%
      \put(3963,484){\makebox(0,0){\strut{}\large 0.2}}%
      \put(4673,484){\makebox(0,0){\strut{}\large 0.25}}%
      \put(5383,484){\makebox(0,0){\strut{}\large 0.3}}%
      \put(6093,484){\makebox(0,0){\strut{}\large 0.35}}%
      \put(6803,484){\makebox(0,0){\strut{}\large 0.4}}%
      \put(286,2739){\rotatebox{-270}{\makebox(0,0){\Large $G_C(Q^2 = 0.16$ GeV$^2$)}}}%
      \put(3962,154){\makebox(0,0){\Large $m_{\pi}^2$ (GeV$^2$)}}%
    \put(0,0){\includegraphics{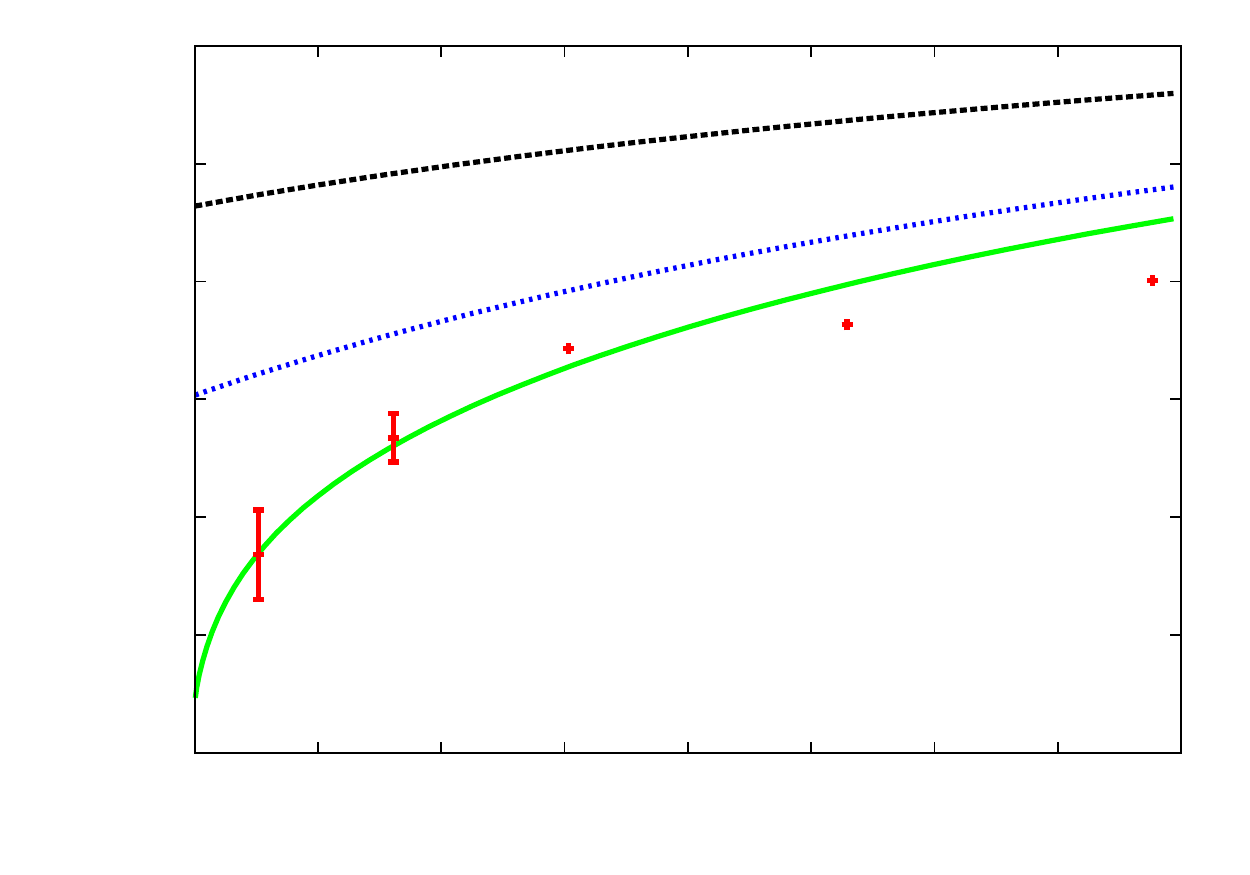}}%
  \end{picture}%
} \\
\centering\resizebox{0.50\textwidth}{!}{\setlength{\unitlength}{0.0500bp}%
  \begin{picture}(7200.00,5040.00)%
      \put(858,704){\makebox(0,0)[r]{\strut{}\large 1.4}}%
      \put(858,1383){\makebox(0,0)[r]{\strut{}\large 1.5}}%
      \put(858,2061){\makebox(0,0)[r]{\strut{}\large 1.6}}%
      \put(858,2740){\makebox(0,0)[r]{\strut{}\large 1.7}}%
      \put(858,3418){\makebox(0,0)[r]{\strut{}\large 1.8}}%
      \put(858,4097){\makebox(0,0)[r]{\strut{}\large 1.9}}%
      \put(858,4775){\makebox(0,0)[r]{\strut{}\large 2}}%
      \put(990,484){\makebox(0,0){\strut{}\large 0}}%
      \put(1717,484){\makebox(0,0){\strut{}\large 0.05}}%
      \put(2443,484){\makebox(0,0){\strut{}\large 0.1}}%
      \put(3170,484){\makebox(0,0){\strut{}\large 0.15}}%
      \put(3897,484){\makebox(0,0){\strut{}\large 0.2}}%
      \put(4623,484){\makebox(0,0){\strut{}\large 0.25}}%
      \put(5350,484){\makebox(0,0){\strut{}\large 0.3}}%
      \put(6076,484){\makebox(0,0){\strut{}\large 0.35}}%
      \put(6803,484){\makebox(0,0){\strut{}\large 0.4}}%
      \put(286,2739){\rotatebox{-270}{\makebox(0,0){\Large $G_M(Q^2 = 0.16$ GeV$^2$)}}}%
      \put(3896,154){\makebox(0,0){\Large $m_{\pi}^2$ (GeV$^2$)}}%
    \put(0,0){\includegraphics{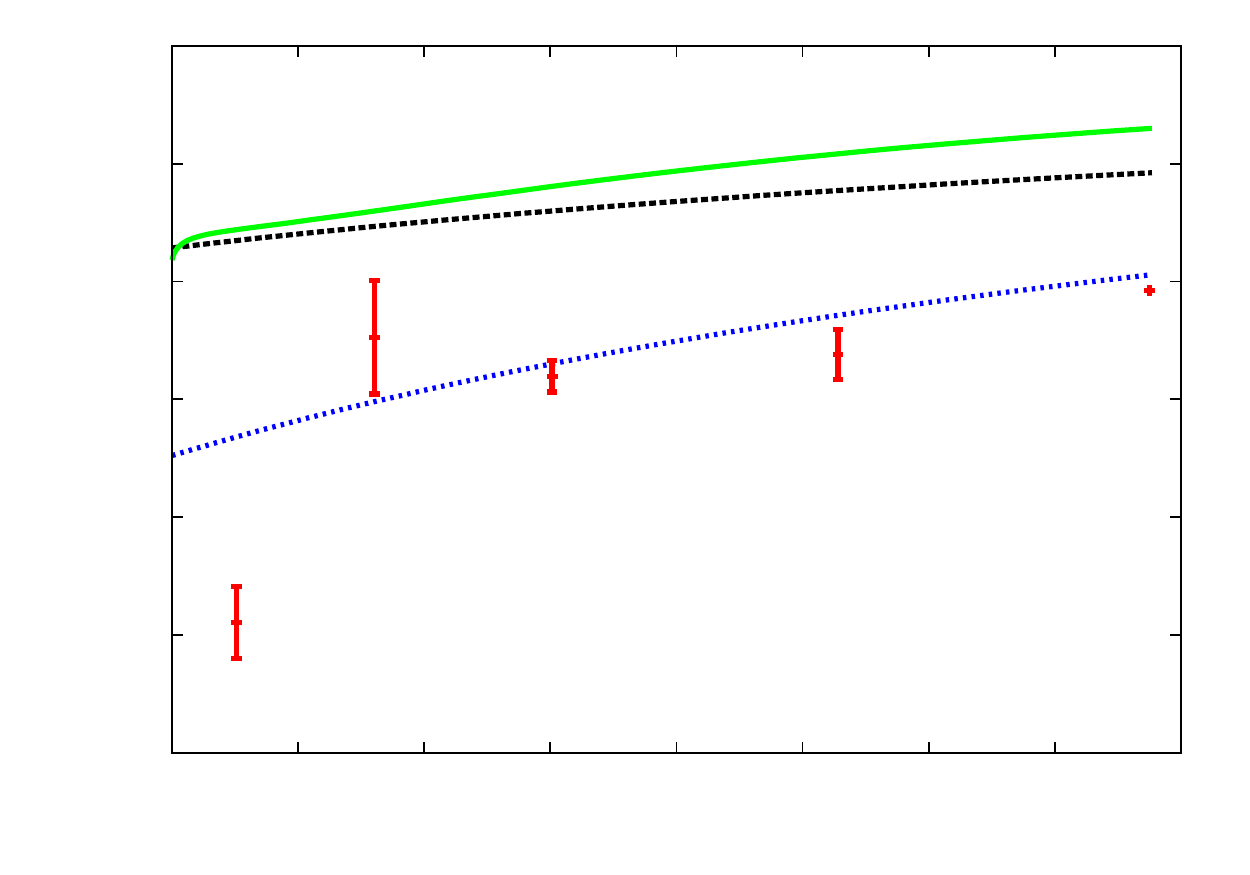}}%
  \end{picture}%
} \\
\centering\resizebox{0.50\textwidth}{!}{\setlength{\unitlength}{0.0500bp}%
  \begin{picture}(7200.00,5040.00)%
      \put(990,704){\makebox(0,0)[r]{\strut{}\large -0.8}}%
      \put(990,1213){\makebox(0,0)[r]{\strut{}\large -0.7}}%
      \put(990,1722){\makebox(0,0)[r]{\strut{}\large -0.6}}%
      \put(990,2231){\makebox(0,0)[r]{\strut{}\large -0.5}}%
      \put(990,2739){\makebox(0,0)[r]{\strut{}\large -0.4}}%
      \put(990,3248){\makebox(0,0)[r]{\strut{}\large -0.3}}%
      \put(990,3757){\makebox(0,0)[r]{\strut{}\large -0.2}}%
      \put(990,4266){\makebox(0,0)[r]{\strut{}\large -0.1}}%
      \put(990,4775){\makebox(0,0)[r]{\strut{}\large 0}}%
      \put(1122,484){\makebox(0,0){\strut{}\large 0}}%
      \put(1832,484){\makebox(0,0){\strut{}\large 0.05}}%
      \put(2542,484){\makebox(0,0){\strut{}\large 0.1}}%
      \put(3252,484){\makebox(0,0){\strut{}\large 0.15}}%
      \put(3963,484){\makebox(0,0){\strut{}\large 0.2}}%
      \put(4673,484){\makebox(0,0){\strut{}\large 0.25}}%
      \put(5383,484){\makebox(0,0){\strut{}\large 0.3}}%
      \put(6093,484){\makebox(0,0){\strut{}\large 0.35}}%
      \put(6803,484){\makebox(0,0){\strut{}\large 0.4}}%
      \put(286,2739){\rotatebox{-270}{\makebox(0,0){\Large $G_Q(Q^2 = 0.16$ GeV$^2$)}}}%
      \put(3962,154){\makebox(0,0){\Large $m_{\pi}^2$ (GeV$^2$)}}%
    \put(0,0){\includegraphics{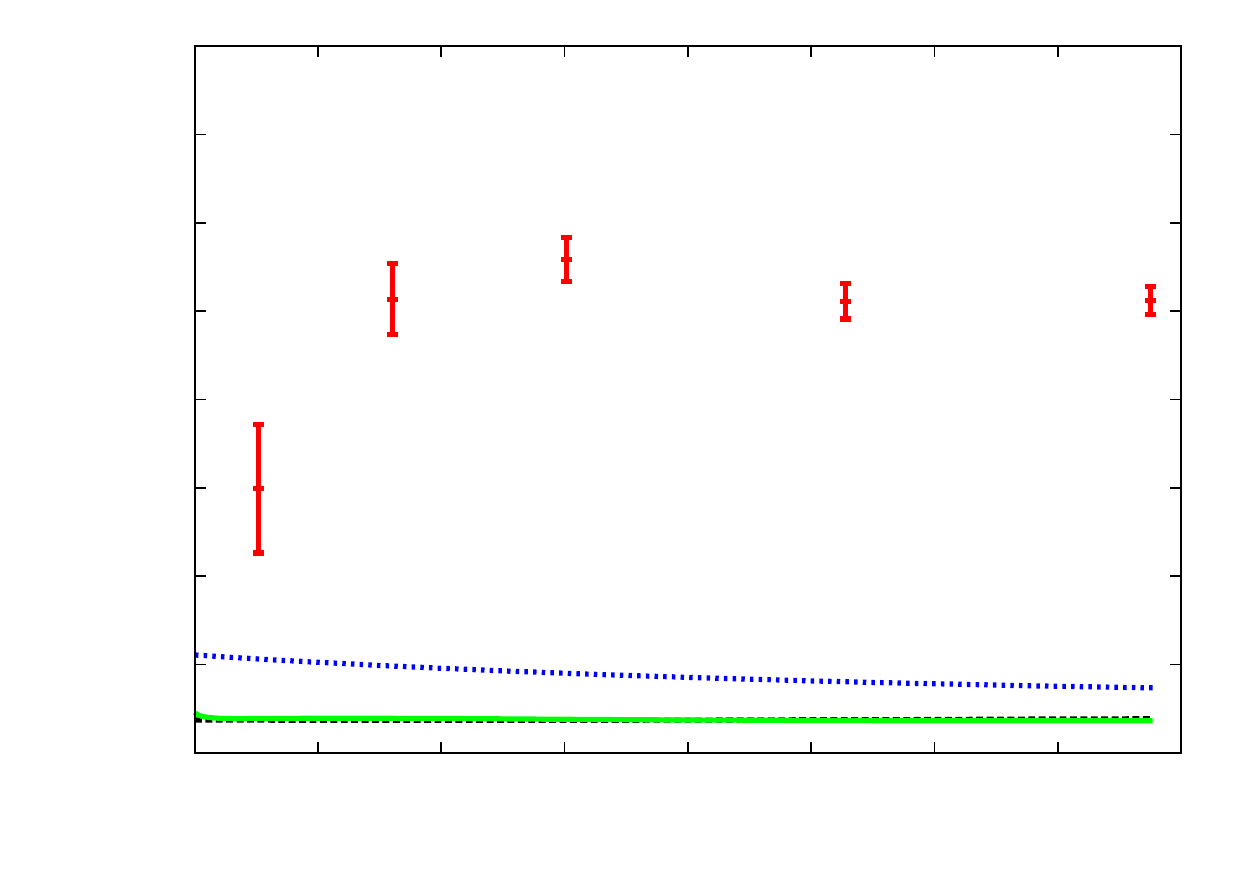}}%
  \end{picture}%
}
\caption{(Colour online) The $\rho^+$ Sachs form factors as a functions of $m_{\pi}^2$ at 
$Q^2 = 0.16$ GeV$^2$. The curves have the same conventions as Fig.~\ref{fig:radius}.}
\label{fig:Gq}
\end{figure}

Finally we find a large quadrupole moment comparable to the Dyson-Schwinger equation
results of Roberts~\textit{et al.}~\cite{Roberts:2011wy} and Hawes~\textit{et al.}~\cite{Hawes:1998bz}.
The lattice QCD result of Ref.~\cite{Owen:2015gva} is approximately $\sim\!30\%$ smaller
than our result near the physics point, as illustrated in Fig.~\ref{fig:Q}. However, as a
hypothesis for the difference we suggest that it may be worthwhile to investigate the
effect of the lack of spherical
symmetry on the lattice simulation, considering that the quadrupole moment reflects
the shape of the $\rho$.

\begin{figure}[tbp]
\centering\resizebox{0.5\textwidth}{!}{\setlength{\unitlength}{0.0500bp}%
  \begin{picture}(7200.00,5040.00)%
      \csname LTb\endcsname%
      \put(990,817){\makebox(0,0)[r]{\strut{}\large -1}}%
      \put(990,1383){\makebox(0,0)[r]{\strut{}\large -0.5}}%
      \put(990,1948){\makebox(0,0)[r]{\strut{}\large 0}}%
      \put(990,2513){\makebox(0,0)[r]{\strut{}\large 0.5}}%
      \put(990,3079){\makebox(0,0)[r]{\strut{}\large 1}}%
      \put(990,3644){\makebox(0,0)[r]{\strut{}\large 1.5}}%
      \put(990,4210){\makebox(0,0)[r]{\strut{}\large 2}}%
      \put(990,4775){\makebox(0,0)[r]{\strut{}\large 2.5}}%
      \put(1122,484){\makebox(0,0){\strut{}\large 0}}%
      \put(2069,484){\makebox(0,0){\strut{}\large 0.2}}%
      \put(3016,484){\makebox(0,0){\strut{}\large 0.4}}%
      \put(3963,484){\makebox(0,0){\strut{}\large 0.6}}%
      \put(4909,484){\makebox(0,0){\strut{}\large 0.8}}%
      \put(5856,484){\makebox(0,0){\strut{}\large 1}}%
      \put(6803,484){\makebox(0,0){\strut{}\large 1.2}}%
      \put(352,2739){\rotatebox{-270}{\makebox(0,0){\Large $G^{\rho}_{i}$($Q^2$) ($m_{\pi}^2 = 0.49$ GeV$^2$)}}}%
      \put(3962,154){\makebox(0,0){\Large $Q^2$ (GeV$^2$)}}%
      \put(1453,3192){\makebox(0,0){\large $G_{C}$}}%
      \put(1453,4549){\makebox(0,0){\large $G_{M}$}}%
      \put(1453,1269){\makebox(0,0){\large $G_{Q}$}}%
    \put(0,0){\includegraphics{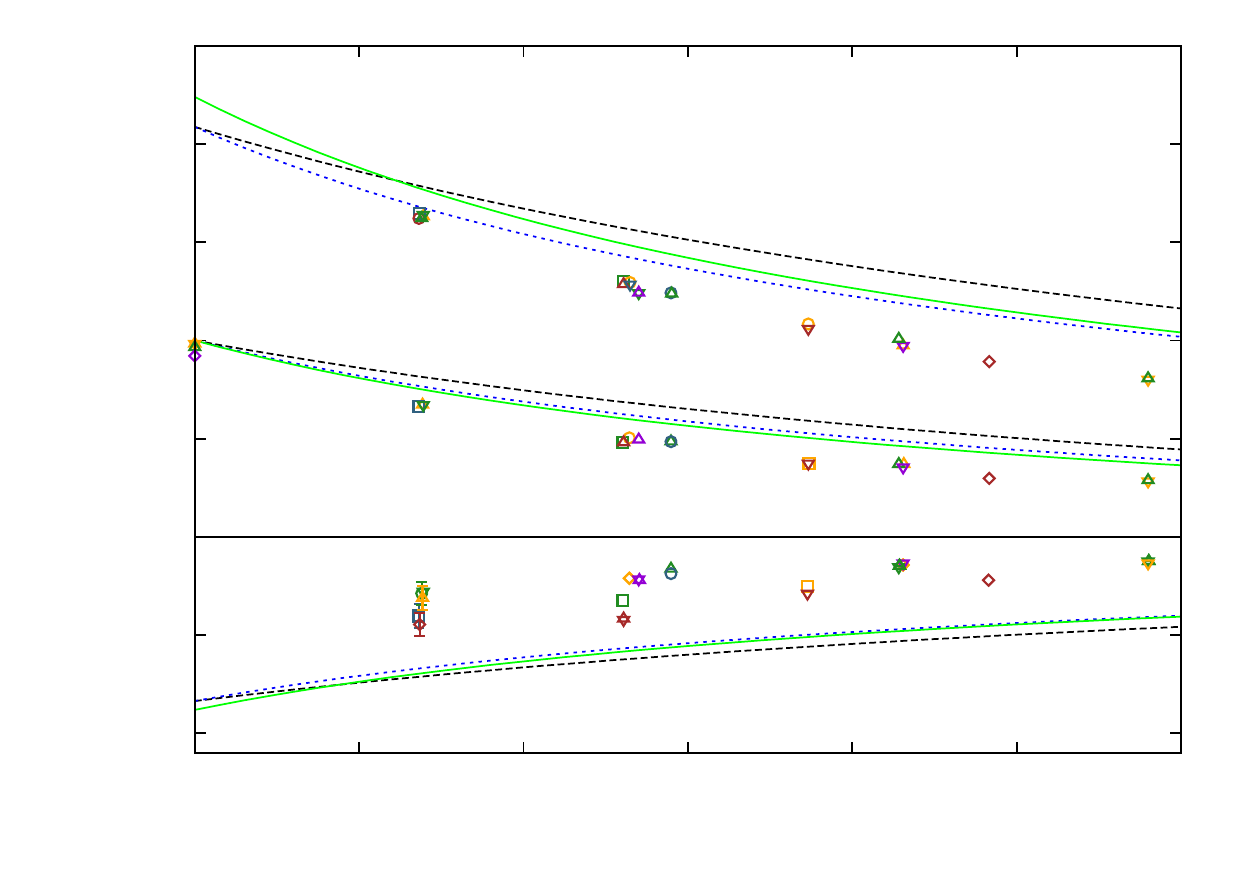}}%
 \end{picture}%
}
\caption{(Colour online) The $\rho^+$ Sachs form factors for $m_{\pi}^2=0.49$ GeV$^2$.
Comparison is made with lattice results from Ref.~\cite{Shultz:2015pfa} and
the curves have the same conventions as Fig.~\ref{fig:radius}.}
\label{fig:ffshultz}
\end{figure}
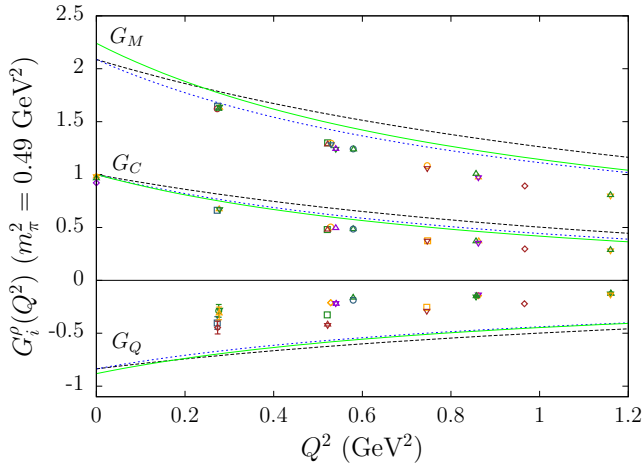

Comparison with the lattice simulation of Ref.~\cite{Owen:2015gva} for the evolution of 
the $\rho^+$ Sachs form factors with $m_{\pi}^2$, at a fixed $Q^2 = 0.16\,$GeV$^2$, 
is made in Figs.~\ref{fig:Gq}. The charge form factor, $G_{C}$, is in good agreement 
with the lattice QCD points, when both the inhomogeneous BSE and pion cloud dressing are included.
On the other hand, for the magnetic form factor $G_{M}$, the BSE results alone
have better agreement with lattice and the pion cloud causes an overestimate. 
The deviations are still small however, considering the simplicity of the calculation. 
The deviation from the lattice simulation data for $G_Q$ is possibly explained 
by the same reason behind the disagreement with $\calQ_\rho$, that is, the lack of spherical
symmetry in the lattice simulation.

A final comparison is made in Fig.~\ref{fig:ffshultz} for the Sachs form factors
as a function of $Q^2$ for a pion mass of $m_{\pi}^2 = 0.49$ GeV$^2$, where the lattice
results are from Ref.~\cite{Shultz:2015pfa}. We find that our model qualitatively describes
the $\rho^+$ form factors obtained from the lattice computation. Once
again the addition of the pion cloud causes an overestimate of $G_M(Q^2)$ and 
the magnitude of $G_Q(Q^2)$ also appears too large.

\section{CONCLUSIONS\label{sec:con}}

We computed the electromagnetic form factors of the $\rho^+$ meson using an NJL model that 
simulates aspects of quark confinement. The quark-photon vertex is studied in three levels 
of sophistication: pointlike dressed quark, via the inhomogeneous BSE and
also including corrections from a pion cloud. The results are qualitatively in good agreement
with the recent lattice QCD computations. 

The main level of disagreement comes from the quadrupole moment and the corresponding form factor. 
We suggest that lattice QCD studies of this type should look at the possible
effects of the lack of spherical symmetry of a cubic lattice in the quadrupole moments and
form factors. It would certainly be helpful to have further lattice studies over a range of 
pion masses and momentum transfers. Experimental measurements would also be extremely valuable.
 
Therefore, the present work on the $\rho$-meson structure and the progress in the computation
of the electromagnetic form factors of the $\pi$ and $K$, including the pion cloud, reported 
in Ref.~\cite{Ninomiya:2014kja}, support the importance of the model as a tool to
describe hadronic structure.
In addition, the NJL model is a quantum field theory where calculations are relatively straightforward 
and it gives good results when compared to more sophisticated methods that require much more resources, 
such as lattice QCD. These
advantages are useful in order to perform larger calculations in problems such as the
description of hadrons in the nuclear medium, as required, for example, to explore the properties of neutron stars.
In such cases the
NJL model serves as a very useful tool to guide possible future computations of lattice QCD and other
more sophisticated approaches.

\begin{acknowledgments}
This work was supported by the Department of Energy, Office of Nuclear Physics, 
contract no. DE-AC02-06CH11357;
the Australian Research Council through the 
ARC Centre of Excellence in Particle Physics at the Terascale 
and an ARC Australian Laureate Fellowship FL0992247 (AWT);
and the Grant in Aid for Scientific Research (Kakenhi) of the Japanese Ministry of
Education, Sports, Science and Technology, Project No. 25400270.
\end{acknowledgments}

\end{document}